\documentstyle[11pt,newpasp,twoside,epsf]{article} 
\def\edcomment#1{\iffalse\marginpar{\raggedright\sl#1\/}\else\relax\fi} 
\def\simless{\mathbin{\lower 3pt\hbox
   {$\rlap{\raise 5pt\hbox{$\char'074$}}\mathchar"7218$}}} 
\def\simgreat{\mathbin{\lower 3pt\hbox
   {$\rlap{\raise 5pt\hbox{$\char'076$}}\mathchar"7218$}}} 
\def\kms{\rm km~s^{-1}}

\def\pc3{\rm pc^{-3}}
\def\erg{\rm erg}
\def\msun{\rm M_\odot}

\def\xte{\rm XTE~J1118+480}

\reversemarginpar 

\begin{document} 
\title{Unveiling black holes ejected from globular clusters}

\author{Alessia Gualandris, Monica Colpi} 
\affil{Dipartimento di Fisica G.Occhialini, 
Universit\`a degli Studi di Milano Bicocca, Piazza della Scienza 3, 
I-20126 Milan, Italy} 
\author{Andrea Possenti} 
\affil{Oss. Astronomico di Bologna, via Ranzani 1, 
I-40127 Bologna, Italy} 
 
\begin{abstract} 
Was the black hole in $\xte$ ejected from a globular cluster
or kicked away from the galactic disk? 
\vskip -1.0truecm
\end{abstract}

\section{The intriguing origin of the kinematic of $\xte$} 
Strong dynamical evidence now exists for 17 black holes (BHs) in X-ray
novae but only a few have been observed away from the galactic plane
(White \& van Paradijs 1996).  $\xte$ is a high galactic latitude
($b=62.3\deg$) BH X-ray nova having an exceptionally large mass
function $~f=6.0 \pm 0.4~\msun$ and the shortest orbital period
($P$=4.08 h) among the BH binaries.  The donor star is probably a main
sequence star of $\sim 0.3~\msun$.  The transverse motion on the plane
of the sky indicates that the system is now orbiting with a space
velocity of $\sim 145~\kms$ relative to the Local Standard of Rest
(Mirabel et al 2001), much higher than the velocity dispersion of the
galactic BH X-ray binaries ($40~\kms$).

{\it The most natural birthplace of the BH in $\xte$ would be the Disk of
the Galaxy, from which a kick could have propelled $\xte$ into the
halo}. This invokes two alternatives.

\noindent
{\bf Disk-A:} An asymmetric natal kick as large ($\sim 200$ km/s)
as never recorded in the case of a black hole. Given the scaling
$V_{{\rm kick,BH}}= V_{\rm {kick,NS}}(m_{\rm NS}/m_{\rm BH})^\alpha$
where $\alpha <1$ (Colpi \& Wasserman 2002), such kick is compatible with
the high velocity end of the observed distribution of the transverse
velocities of the pulsars (Arzoumanian, Chernoff  \& Cordes 2002).

\noindent
{\bf Disk-B:} A symmetric kick from recoil due to supernova
explosion (typically of order $\sim 40$ km/s), superimposed to the
contribution of the circular velocity ($> 100$ km/s) of the galactic
disk at the site of the supernova explosion.

\noindent
In Figure 1 we show the orbit followed by $\xte$ during the last 3
Gyrs; we have verified that some of the transits across the galactic
disk are cinematically consistent with the hypothesis {\bf Disk-B}.

\begin{figure}
\plottwo{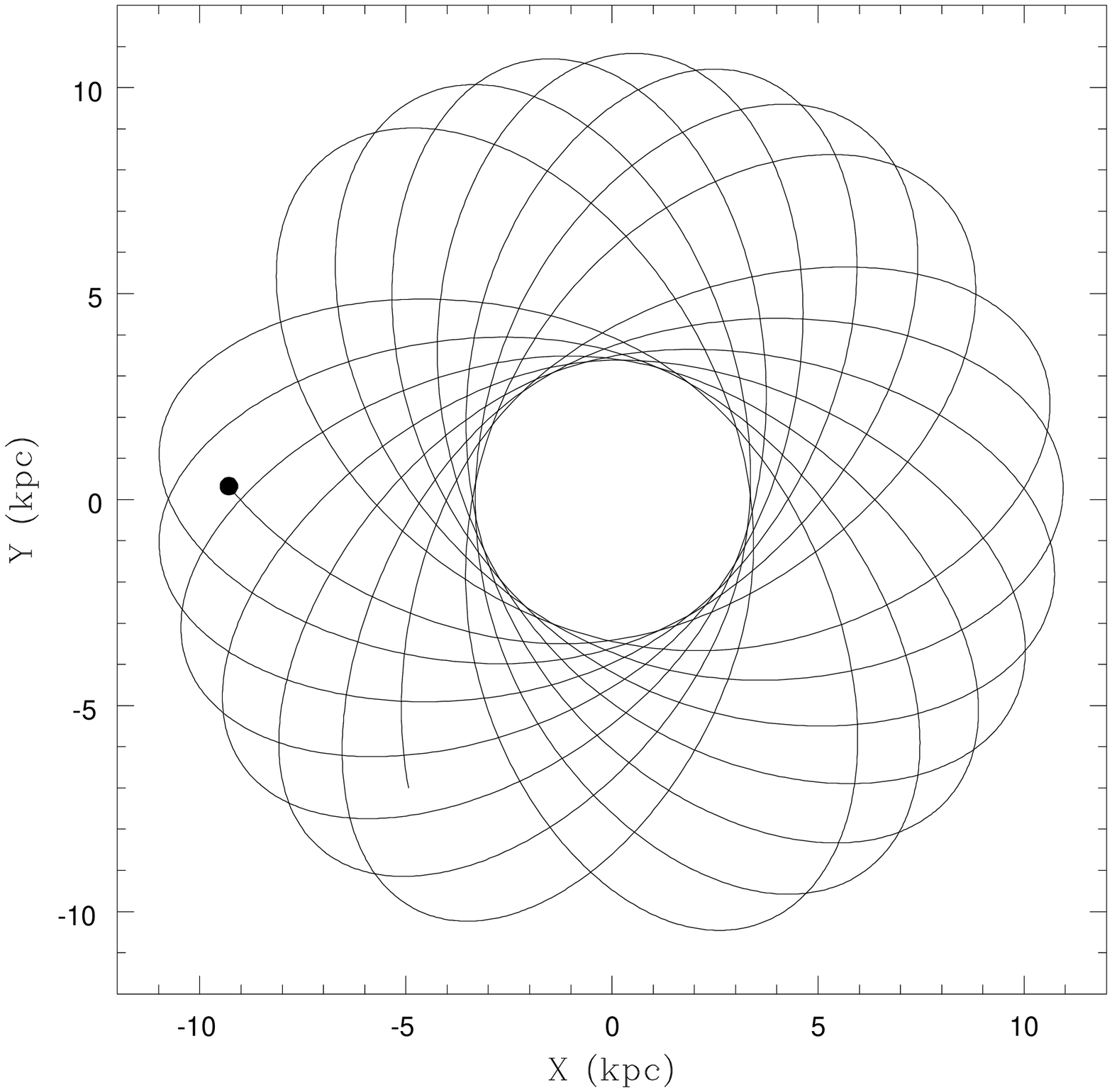}{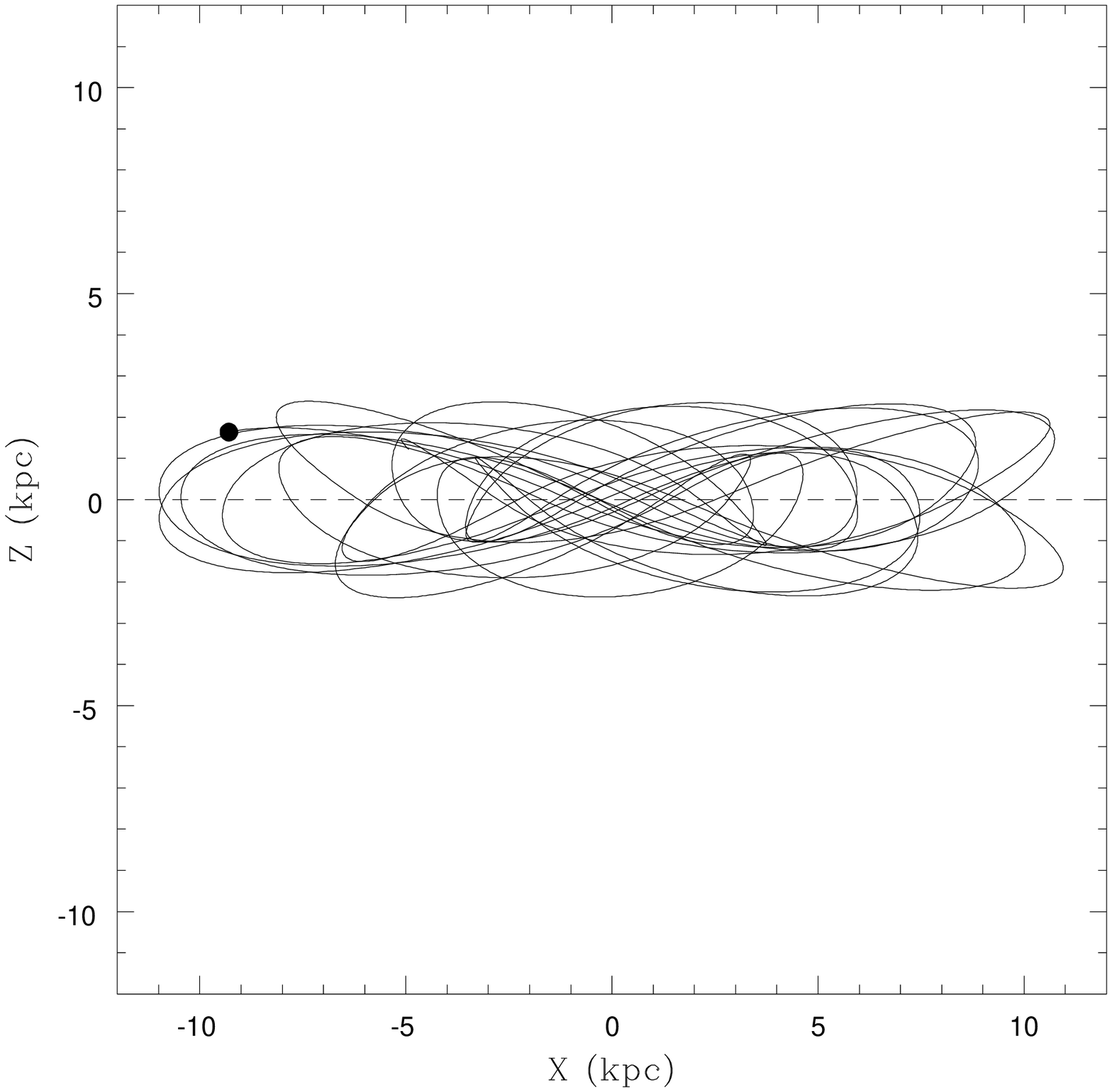}
\vskip -0.1truecm
\caption{Galactic orbit (solid line) and current position (full dot) of $\xte$ 
integrated backward in time for 3 Gyr.} 
\vskip -0.1truecm
\end{figure}

{\it A completely different scenario calls for $\xte$ to be ejected from a
globular cluster} (GC). Two expulsion mechanisms have been explored.

\noindent
{\bf GC-A:} A three-body exchange interaction (where a single
BH intrudes into a main-sequence star binary, MS+ms, replacing the
less massive star, ms) occurred in the cluster core in the first
stages of its evolution.  However, given the recoil energy of $\sim
10^{47}~\erg$ needed to eject the BH+MS binary from a GC of typical
escape speed 30-50 $\kms,$ the initial MS+ms binary should have been
extremely tight (at most $3-4~{\rm R_\odot}$) for accomplishing both
momentum and energetic requirements.  As a consequence, the rate for
this interaction is extremely small, having a time-scale $T \approx
10^{12}$ yr (for a stellar density $n\sim 10^4~\pc3$ and a BH
fraction of $\sim 2\%$). Furthermore, such a close approach between
the BH and the other bodies  
can trigger tidal effects and mass loss in any of the main
sequence stars leading to angular momentum losses and hence lower
recoil velocities for the MS+BH binary.

\noindent
{\bf GC-B:} A kick imparted to the binary at the moment of the
BH formation due to the recoil from the SN explosion.  The recoil
velocity required for escaping the GC potential well ($V_{\rm rec}\sim
40-50~\kms$) is compatible with the velocity dispersion observed in
the sample of the BH X-ray binaries in the galactic field and can be
achieved with the ejection of only $\sim 10~\msun$, which in turn
allows to avoid the disruption of the massive ($M_{\rm tot}>25 {\rm
M_\odot}$) pre-SN binary system.

The morphological similarity between the galactic orbit of $\xte$ and
that of some GCs suggested (Mirabel et al. 2001) the fascinating
opportunity of identifying the parent cluster of $\xte$ using purely
kinematic considerations, namely the coincidence in the 6-dimensional
phase space (of velocities and positions) at a given time.
Unfortunately, we have verified that the current uncertainties in the
observed velocity vector of the GCs (having errors as large as $\sim
20\%$, Dinescu et al. 1999) hamper any reliable identification if the
ejection has occurred (like in the cases {\bf GC-A} and {\bf GC-B})
earlier than the last $\sim$ 2-3 Gyr.

\end{document}